# Anomalous domain periodicity observed in ferroelectric PbTiO$_3$ nanodots with 180$^\text{o}$ stripe domains

**Jong Yeog Son,**[1] **Jung-Hoon Lee,**[2] **Young-Han Shin,**[3] **and Hyun Myung Jang**[2,4]*

[1]Department of Applied Physics, College of Applied Science, Kyung Hee University, Suwon 446-701, Republic of Korea

[2]Department of Materials Science and Engineering and Division of Advanced Materials Science, Pohang University of Science and Technology (POSTECH), Pohang 790-784, Republic of Korea

[3]Department of Physics and EHSRC, University of Ulsan, Ulsan 680-749, Republic of Korea

[4]Department of Physics, Pohang University of Science and Technology (POSTECH), Pohang 790-784, Republic of Korea

E-mail: hmjang@postech.ac.kr

**Abstract.** PbTiO$_3$-based nano-scale dots and tubes have received a great deal of attention owing to their potential applications to nonvolatile memories and multi-functional devices. As for the size effect of 180$^\text{o}$ stripe domains in ferroelectric thin films, there have been extensive reports on the thickness-dependent domain periodicity. All these studies have revealed that the domain periodicity of 180$^\text{o}$ stripe domains scales with the film thickness ($d$) according to the classical Landau-Lifshitz-Kittel (LLK) scaling law down to the thickness of ~2 *nm*. In the case of PbTiO$_3$ nanodots, however, we obtained a quite striking correlation that for the thickness less than a certain critical value, $d_c$ (~35 *nm*), the domain width even increases with decreasing thickness of the nanodot, which surprisingly indicates a negative value in the LLK scaling-law exponent. On the basis of theoretical considerations of $d_c$, we attribute this anomalous domain periodicity to the presence of a nonferroelectric surface layer, in addition to the finite lateral-size effect of a ferroelectric nanodot.

\* Author to whom any correspondence should be addressed.

**Contents**

# 1. Introduction

Ferroelectrics are receiving a great deal of attention because of their technological promise in leading toward miniaturized and efficient memory devices [1]. Like other ferroics, ferroelectrics are characterized by domain structures. Various forms of ferroelectric materials, such as ceramics, single crystals and thin films, exhibit a variety of different domain structures which include a stripe, mosaic, or vortex to minimize the total free energy which is composed of competing depolarization and domain-wall energy terms [2-10].

Ferroic domains that are ordered along one unique direction but with opposite polarity or magnetic moment are called $180^\circ$ domains. It has long been known that the width of $180^\circ$ stripe magnetic domains closely follows the so-called Landau-Lifshitz-Kittel (LLK) scaling law [11]. This law was later extended to ferroelectric domains by Mitsui and Furuichi [2]. According to the scaling law, the domain width ($w$) is directly proportional to the square root of the crystal thickness ($d$), namely, $w = Ad^\gamma$, where $A$ is a proportionality constant and $\gamma$ is the scaling-law exponent (= 1/2).

Till now, extensive theoretical studies [3,12-16] have been carried out to examine the validity of the LLK scaling law on ferroelectric multiple domains, in conjunction with a variety of experimental studies done by employing various methods that include x-ray scattering [5,6], piezoelectric force microscopy [10], and scanning transmission electron microscopy [12]. All these studies reveal that the LLK scaling law with the exponent around 1/2 is valid down to the thickness of ~2 $nm$ [5,14,15], except for one interesting study reported by Catalan *et al* [10]. According to their study, the domain size of multiferroic BiFeO$_3$ thin films having irregular domain walls is substantially larger than those of other ferroelectrics having the same thickness and the observed scaling-law exponent ($\gamma$) of 0.59 deviates quite substantially from its normal value of 1/2. They correlated the former with a strong magnetoelectric coupling at domain walls while attributing the latter to a fractal-like Hausdorff dimension [10].

Among numerous ferroelectrics, lead titanate (PbTiO$_3$; PTO hereafter) has been most extensively studied and is known as a prototype of robust displacive ferroelectrics without exhibiting any over-damping of the resonance-type soft phonons [17,18]. Currently, PTO-based nano-scale dots and tubes have received a great deal of attention owing to their potential applications to high-density nonvolatile memories and multi-functional devices [19,20]. As for the size effect of 180$^\circ$ stripe domains in PTO thin films, there have been extensive reports on the thickness-dependent domain periodicity [5,6,13-15]. All these studies reveal that the domain periodicity of 180$^\circ$ stripe domains scales with the film thickness ($d$) according to the classical LLK scaling law down to the thickness of ~2 *nm* [5,14,15]. On the contrary, a theoretical solution of the Laplace equation for rigorously evaluating the depolarizing-field energy suggests that the LLK scaling law does break for the thickness less than a certain critical size, $d_c$ [21]**:** the domain width ($w$) even increases with decreasing thickness below $d_c$.

In view of the above discrepancy between the theoretical prediction [21] and the experimental observations [5,14,15], it is of great scientific importance to experimentally clarify whether there exists any critical thickness ($d_c$) below which the LLK scaling law is no more valid. Until now, however, the validity of the LLK scaling law for 180$^\circ$ stripe ferroelectric domains has been experimentally tested using thin films [5,6,10,12-15] where the lateral dimension ($L$) is practically infinite. Considering this, we have critically examined the effect of the lateral dimension on the validity of the LLK scaling law using ferroelectric PTO nanodots having a variety of different lateral sizes.

Herein we present a quite striking experimental result that the thickness-dependent domain width of the PTO nanodot does not obey the LLK scaling law but is characterized by a negative exponent ($\gamma < 0$) below a certain critical thickness, $d_c$. On the basis of theoretical considerations of $d_c$, we have concluded that the existence of a ferroelectrically inactive surface layer, in addition to the finite lateral-size effect, should be taken into account to properly explain the difference in the scaling behavior between ferroelectric films and dots.

## 2. PbTiO$_3$ nanodots with ferroelectric 180$^\circ$ stripe domains

A ferroelectric PTO nanodot array was fabricated on a Nb-doped SrTiO$_3$ (STO) substrate using the dip-pen nanolithography (DPN) method [figure 1(a)]. This method had been successfully used in the fabrication of PTO nanodots having a variety of different lateral sizes [20]. Figure 1(b) shows atomic force microscopy (AFM) images of PTO nanodots with several different size classes. These DPN-formed nanodots are rectangle-shaped, indicating a high degree of crystallinity. From the high-resolution transmission electron microscopy image, we confirmed that these PTO nanodots are characterized by the tetragonal *4mm* ($C_{4v}$) symmetry with the polar *c*-axis perpendicular to the substrate plane [20].

As shown in figure 1(c), the dimension (thickness, side length) of the nanodot can be controlled by adjusting the dip-pen deposition time. For a short deposition time (< 1 sec), the lateral dimension (side length) increases rapidly while the thickness increases rather steadily with increasing deposition time. For the four PTO nanodots labeled A, B, C, and D [figure 1(b)], piezoelectric hysteresis loops were measured by employing piezoelectric force microscopy (PFM) at a frequency of 10 kHz. As shown in figure 1(d), $d_{33}$ value decreases while the coercive electric field increases with the lateral size, which presumably reflects size-dependent depolarization effects [20].

To examine the effect of the lateral dimension on the validity of the LLK scaling law, PFM measurements were carried out for a series of PTO nanodots having a variety of different lateral sizes ranging from 45 nm to 500 nm. PFM images were observed using a high-resolution electric force mode of the PFM, where a platinum-coated Si$_3$N$_4$ cantilever tip was employed. Figure 2(a) presents PFM images of the five selected PTO nanodots with different lateral sizes. The PFM images indicate that regardless of the lateral dimension, the PTO nanodots grown on a Nb-doped STO substrate are characterized by ferroelectric 180$^o$ stripe domains.

The PFM line profile of the PTO nanodot having a lateral dimension of 185 nm is shown in figure 2(b) as an example. This line profile demonstrates that the nanodot is composed of nine 180$^o$ stripe domains. Figure 2(c) presents the AFM line profiles of the four selected nanodots [A, B, C, and D; figure 1(b)] showing that the thickness of the nanodot increases with the lateral dimension. Similar to the present nanodots, 180° stripe-domain structures were also observed in epitaxially grown PTO films (on STO substrates) for the film thickness down to ~2 nm [5].

## 3. Anomalous domain periodicity observed in PbTiO$_3$ nanodots

We now focus on the correlation between the domain width ($w$) and the dot thickness ($d$). In the case of PTO thin films, there is a good linear correlation between $\log w$ and $\log d$ with the scaling exponent ($\gamma$) of 1/2 [figure 3], which closely follows the LLK scaling law of $w = A d^{\gamma}$ [5]. On the contrary, the PTO nanodots exhibit a quite striking correlation. For $d > 35$ nm, the PTO nanodots also follow the scaling law with the estimated exponent of 0.52. However, the domain width even increases with decreasing thickness for $d < 35$ nm [figure 3]. This kind of surprising results has never been observed in ferroelectric thin films.

To clarify the main cause of the striking result observed in the PTO nanodots, we have theoretically considered the effect of the lateral dimension on the difference in the scaling behavior between a nanodot and a thin film and consequently examined the possibility of occurrence of the anomalous domain periodicity for the dot thickness less than a certain critical value ($d_c$). For this purpose, we first formulated the Gibbs free-energy of a ferroelectric nanodot (having a finite lateral dimension) as a function of the domain width and obtained a modified LLK scaling law that accounts for, at least qualitatively, the observed anomalous domain periodicity for $d < 35$ nm.

## 4. Gibbs free-energy function of a ferroelectric nanodot

Let us consider a ferroelectric nanodot (*i.e.*, nano-rectangle) having a dimension of $L*L*d$ with the domain width of $w$, as schematically depicted in figure 4. Then, the Gibbs free-energy function of a ferroelectric nanodot with respect to that of a paraelectric nanodot (at given $T$ and $P$) can be written in terms of $w, d,$ and $L$ as

$$\Delta G^{dot} = -\Delta\mu\left(L^2 d\right) + \Delta G_{dep}(L^2) + \sigma_w\left(\frac{L}{w}-1\right)Ld + \sigma_s(4Ld + 2L^2) \tag{1}$$

where $\Delta\mu$ denotes the difference in the bulk free energy between the paraelectric and ferroelectric phases at given $T$ and $P$, namely, $\Delta\mu \equiv G^o_{para} - G^o_{ferro} > 0$ and $\Delta G_{dep}$ designates the depolarization-field energy per unit area. On the other hand, $\sigma_w$ in equation (1) denotes the domain-wall energy per unit area whereas $\sigma_s$ represents the surface tension of four side faces of a retangular nanodot having the area of $L*d$ per face. Here $\sigma_s$ can be viewed as the excess surface free energy (per unit area) of the ferroelectric rectangle $[\sigma(T,P)]$ with respect to the surface tension of the paraelectric rectangle ($\sigma_p$) having the same dimension, *i.e.*, $\sigma_s \equiv \sigma(T,P) - \sigma_p$. Thus, the last term takes care of the excess surface free energy of a nano-rectangle having six mutually orthogonal faces. For simplicity, we assume that $\sigma_s = \sigma'_s$, where $\sigma'_s$ denotes the excess surface free energy of top (or bottom) surface of a retangular nanodot.

According to the Landau-Ginzburg theory, the difference in the free energy between ferroelectric and paraelectric states can be expanded in terms of $P$ (polarization order-parameter) and its gradient as

$$\begin{aligned} G_{ferro} - G_{para} &= (G^o_{ferro} - G^o_{para}) + \frac{\kappa}{2}|\nabla \mathbf{P}|^2 \equiv -\Delta\mu + \frac{\kappa}{2}|\nabla \mathbf{P}|^2 \\ &= \frac{1}{2}\chi P^2 + \frac{1}{4}\xi P^4 + \frac{1}{6}\zeta P^6 + \frac{\kappa}{2}|\nabla \mathbf{P}|^2 \end{aligned} \tag{2}$$

where $\kappa$ denotes the Ginzburg gradient-energy coefficient. In equation (2), $\chi$ (below the Curie temperature) and $\xi$ are negative while $\zeta$ is positive for a displacive ferroelectric that undergoes a discontinuous first-order phase transition [22]. From equation (2), one can deduce the following expression for the equilibrium bulk polarization $\left(P_o^2\right)$ under the condition of zero gradient:

$$P_o^2 = \frac{|\xi| + \sqrt{\xi^2 - 4\chi\zeta}}{2\zeta} \tag{3}$$

Substituting equation (3) into equation (2) yields the following expression that relates $\Delta\mu$ with the Landau expansion coefficients (*i.e.,* dielectric stiffness coefficients):

$$\Delta\mu = \frac{1}{24\zeta^2}\left\{|\xi|\left(|\xi|^2 + 6|\chi|\zeta\right) + \left(|\xi|^2 + 4|\chi|\zeta\right)^{3/2}\right\} > 0 \qquad (4)$$

Considering equation (2), one can evaluate $\sigma_w$ by carrying out the following integration if the polarization profile across the domain wall, $P(x)$, is well established (*e.g.,* $P(x) = P_o \tanh(x/l_w)$, where $l_w$ denotes the half width of the domain wall) [3,14]:

$$\begin{aligned}
\sigma_w &= \int_{-\infty}^{+\infty}\left\{\frac{\chi}{2}(P^2 - P_o^2) + \frac{\xi}{4}(P^4 - P_o^4) + \frac{\zeta}{6}(P^6 - P_o^6) + \frac{\kappa}{2}\left|\frac{\partial \mathbf{P}}{\partial x}\right|^2\right\}dx \\
&= 2\cdot\int_0^{+\infty}\left\{\frac{\chi}{2}(P^2 - P_o^2) + \frac{\xi}{4}(P^4 - P_o^4) + \frac{\zeta}{6}(P^6 - P_o^6) + \frac{\kappa}{2}\left|\frac{\partial \mathbf{P}}{\partial x}\right|^2\right\}dx \approx 2\sigma_s
\end{aligned} \qquad (5)$$

The second expression of equation (5) is valid for a symmetrical domain boundary.

Let us now return to equation (1). Substituting $nw$ for $L$ and dividing $\Delta G^{dot}$ by $L^2$, one obtains the following expression for the Gibbs free-energy of a ferroelectric nanodot per unit area ($\Delta G$):

$$\begin{aligned}
\Delta G &\equiv \frac{\Delta G^{dot}}{L^2} = \Delta G_{sd} + \Delta G_{dep} + \Delta G_w \\
&= -\Delta\mu d + 2\sigma_s\left(\frac{2d}{nw} + 1\right) + \Delta G_{dep} + \frac{(n-1)\sigma_w}{n}\left(\frac{d}{w}\right)
\end{aligned} \qquad (6)$$

where the first two terms in the right-hand-side of equation (6) represent the free energy of the single-domain state per unit area ($\Delta G_{sd}$) whereas the last term denotes the domain-wall energy ($\Delta G_w$) of a nano-rectangle having the dimension of $L*L*d$. $\Delta G_{dep}$ in equation (6) represents the depolarizing-field energy. Let us now define the following parameter for future convenience [21]:

$$R \equiv \frac{\pi c}{2}\left(\frac{d}{w}\right) = \frac{\pi d}{2w}\sqrt{\frac{\varepsilon_x}{\varepsilon_z}}, \quad \text{where} \quad c \equiv \sqrt{\frac{\varepsilon_x}{\varepsilon_z}} \qquad (7)$$

$\varepsilon_z$ in equation (7) denotes the relative dielectric permittivity along the unique polarization axis, *i.e.*, $\varepsilon_c$ for PTO. Then, $\Delta G_w$ can be rewritten in terms of $R$ and $c$ as

$$\Delta G_w = \frac{(n-1)\sigma_w}{n}\left(\frac{d}{w}\right) = \frac{(n-1)}{n}\left(\frac{2R}{\pi c}\right)\sigma_w \tag{8}$$

Similarly, $\Delta G_{sd}$ can be rewritten using $R$ and $c$ as

$$\Delta G_{sd} = -\Delta \mu d + 2\sigma_s \left(\frac{4R}{n\pi c} + 1\right) \tag{9}$$

On the other hand, the following complicated expression of $\Delta G_{dep}$ can be obtained by solving the Laplace equation under suitable continuity conditions for the electric field (**E**) and the dielectric displacement vector (**D**) [21]:

$$\begin{aligned}\Delta G_{dep} &= \frac{P_o^2 d}{2\varepsilon_o \varepsilon_z}\left[\left(\frac{8g}{\pi^2 R}\right)\sum_{m=1,2,\ldots}^{\infty}\frac{1}{m^3}\sin^2\left(\frac{m\pi}{2}\right)\frac{1}{\{1+g\coth(mR)\}}\right] \\ &\equiv \frac{P_o^2 d}{2\varepsilon_o \varepsilon_z} f(R,g)\end{aligned} \tag{10}$$

where $g$ is defined by $g \equiv \sqrt{\varepsilon_x \varepsilon_z} = c\varepsilon_z$. The term, $P_o^2 d / 2\varepsilon_o \varepsilon_z$, represents the energy of a plate condenser per unit area having thickness $d$, filled by a dielectric with the permittivity $\varepsilon_z$ and carrying surface charges $\pm P_o$. On the other hand, $f(R,g)$ is a dimensionless function which is equal to the terms inside the parenthesis of equation (10).

According to Kopal, Bahnik, and Fousek [21], there exists a certain critical film thickness above which the electrostatic interaction of the domain surfaces can be neglected. This thickness is given by

$$d_{c(film)} = 5\pi \varepsilon_o \left(\frac{\varepsilon_z^{3/2}}{\varepsilon_x^{1/2}}\right)\left(\frac{\sigma_w}{P_o^2}\right) \tag{11}$$

For $d > d_c$, $\Delta G_{dep}$ can be simplified by the following well known expression [2,14,21,23], instead of the complicated expression presented in equation (10):

$$\Delta G_{dep} = \frac{7\zeta(3)}{\pi^3} \frac{P_o^2 w}{\varepsilon_o \left(1+\sqrt{\varepsilon_x \varepsilon_z}\right)} = \frac{8.42}{\pi^3} \frac{P_o^2 w}{\varepsilon_o \left(1+\sqrt{\varepsilon_x \varepsilon_z}\right)} \equiv \alpha P_o^2 w \qquad (12)$$

where $\zeta(3)$ is Riemanns zeta function and is equal to $\sum_{n=-1}^{\infty} n^{-3} \approx 1.202$.

## 5. A modified scaling law for ferroelectric nanodots

Under the condition of $d > d_c$ (*i.e.*, "thick" plates approximation), equation (6) can be written explicitly using equation (12) as

$$\Delta G = -\Delta\mu d + 2\sigma_s \left(\frac{2d}{nw}+1\right) + \alpha P_o^2 w + \frac{(n-1)\sigma_w}{n}\left(\frac{d}{w}\right) \qquad (13)$$

Then, one can obtain the following expression for the equilibrium domain width of a ferroelectric rectangle (dot) by setting $\left(\partial \Delta G/\partial w\right)_d = 0$:

$$\left(w_{eq}\right)^2 = \frac{d}{\alpha P_o^2}\left\{\left(\frac{n-1}{n}\right)\sigma_w + \frac{4\sigma_s}{n}\right\} \approx \frac{d}{\alpha P_o^2}\sigma_w \qquad (14)$$

The last expression of equation (14) represents an asymptotic scaling law under the condition of $n \to \infty$ and, thus, corresponds to a thin film having an infinite lateral dimension. As expected, this asymptotic law correctly reproduces the classical $w^2-d$ scaling law. However, the first expression of equation (14) indicates that the $w^2-d$ scaling law may not be valid for a nanodot where $n$ is a small integer and depends on $d$.

One cannot neglect the mutual electrostatic interaction between the domain (plate) surfaces for the dot thickness smaller than $d_c$. In this case, equation (12) should not be used to evaluate $\Delta G_{dep}$. Instead, one can derive a modified scaling law by exploiting equations (8), (9), and (10) and subsequently by setting $(\partial \Delta G/\partial R)_g = 0$. To do this, let us first evaluate $\left(\partial \Delta G_{dep}/\partial R\right)_g$. In doing this, one has to consider the following two obvious relations: $\sin^2(m\pi/2)=1$ for $m$ = 1, 3, 5 …. (odd integers) and $\sin^2(m\pi/2) = 0$ for $m$ = 2, 4, 6 …. (even integers). Incorporating this result into equation (10), one obtains the following expression of $(\partial \Delta G_{dep}/\partial R)_g$:

$$\left(\frac{P_o^2 d}{2\varepsilon_o \varepsilon_z}\right)\left[\left(-\frac{8g}{\pi^2 R^2}\right)\sum_{m(odd)}^{\infty}\frac{1}{m^3\{1+g\cot h(mR)\}} + \left(\frac{8g^2}{\pi^2 R}\right)\sum_{m(odd)}^{\infty}\frac{1}{m^3\{\sin h(mR)+g\cos h(mR)\}^2}\right] \quad (15)$$

Substituting this equation into the requirement that $\partial\{\Delta G_{sd} + \Delta G_w + \Delta G_{dep}\}/\partial R = 0$, one can eventually obtain the following non-classical relation between $w$ and $d$ for $d < d_c$:

$$\left\{\frac{4\sigma_s}{n} + \left(\frac{n-1}{n}\right)\sigma_w\right\}\left(\frac{\varepsilon_o \varepsilon_z \pi}{2\varepsilon_x P_o^2}\right)\left(\frac{1}{d}\right) = h(R) \approx \frac{\varepsilon_o \varepsilon_z \pi}{2\varepsilon_x P_o^2}\sigma_w\left(\frac{1}{d}\right) \quad (16)$$

where

$$h(R) \equiv \left[\frac{1}{R^2}\sum_{m(odd)}^{\infty}\frac{1}{m^3\{1+g\cot h(mR)\}} - \frac{g}{R}\sum_{m(odd)}^{\infty}\frac{1}{m^3\{\sin h(mR)+g\cos h(mR)\}^2}\right] \quad (17)$$

Equation (16) clearly indicates that the simple $w^2 - d$ scaling law is no more valid for $d < d_c$. According to equation (16), the domain width even increases with decreasing $d$ [21], which successfully accounts for the observed anomalous domain periodicity [figure 3] for $d <$ 35 nm ($d_c$). It is interesting to note that the last expression of equation (16) which is asymptotically valid for $n = \infty$ (infinite lateral dimension) exactly coincides with the modified scaling equation proposed previously for thin films [21].

**6. Effect of the lateral dimension on the critical thickness**

Though the anomalous behavior of the domain periodicity (for $d < d_c$) can be explained by adopting equation (16), we still have one important question to be resolved in the case of thin films**:** Why does the classical $w^2 - d$ scaling law describe the domain periodicity well down to the thickness of ~2 *nm* ? In other words, why is the critical thickness ($d_c$) not observed down to ~2 *nm* in the case of thin films ? To answer this question, we have considered the most prominent difference between a nanodot and a thin film, which is the lateral dimension ($L$), thus, the number of domains, $n$.

In the case of thin films where the lateral dimension ($L$) is practically infinite, the critical thickness ($d_c$) for neglecting the electrostatic interaction of the domain surfaces is given by equation (11) [21]. We will then deduce the corresponding expression of the critical thickness for a ferroelectric nanodot ($d_{c(dot)}$) where $L$ or $n$ is finite. To do this, we first assume that the domain width at $d_c$ is proportional to $d_{c(dot)}$ itself. Thus, one can establish that $w_{c(dot)} = k\, d_{c(dot)}$, where $k$ is a proportionality constant. By exploiting this proportionality and equation (14) in the vicinity of $d_c$, one can eliminate $w_{c(dot)}$ from this relation and obtain the following expression for $d_{c(dot)}$:

$$d_{c(dot)} = \frac{1}{\alpha P_o^2 k^2} \left\{ \left(\frac{n_c - 1}{n_c}\right) \sigma_w + \frac{4\sigma_s}{n_c} \right\} = \frac{5\pi \varepsilon_o}{P_o^2} \left(\frac{\varepsilon_z^{3/2}}{\varepsilon_x^{1/2}}\right) \left\{ \left(\frac{n_c - 1}{n_c}\right) \sigma_w + \frac{4\sigma_s}{n_c} \right\}$$
$$\approx \frac{5\pi \varepsilon_o \sigma_w}{P_o^2} \left(\frac{\varepsilon_z^{3/2}}{\varepsilon_x^{1/2}}\right) \qquad (18)$$

where $n_c$ denotes the number of distinct domains in the nanodot having the critical thickness, $d_{c(dot)}$. $k^2$-term appeared in the first expression of the above equation was eliminated in the second expression by comparing equation (11) with the first expression in the asymptotic thin-film limit where $n \to \infty$. It is worth noting that the last expression of equation (18) corresponding to the asymptotic thin-film limit does coincide with equation (11) which had been deduced by Kopal and co-workers for thin films [21].

Comparing the second expression of equation (18) with equation (11), one obtains the following difference in the critical thickness between a nanodot having $n$ distinct 180° domains and a thin-film having an infinite lateral dimension:

$$R_0 \equiv \frac{d_{c(dot)}}{d_{c(film)}} = \left(\frac{n_c - 1}{n_c}\right) + \frac{4\sigma_s}{n_c \sigma_w} \geq 1 \qquad (19)$$

This equation predicts that the asymptotic value of the above ratio in the limit of $n_c \to \infty$ (i.e., thin film) is 1 as expected. One can qualitatively estimate $d_{c(film)}$ using equation (19). Taking $(\sigma_s/\sigma_w) \approx 1/2$, as used in equation (5), and plugging $n_c = 3$ and $d_{c(dot)} \approx 35\ nm$ into equation (19), one predicts that $d_{c(film)} = (3/4) * d_{c(dot)} \approx 26$ *nm* which is in direct disagreement with the observation that $d_{c(film)} < 2$ *nm* [5,14,15]. This suggests that $d_c$ is actually determined by some factor other than the lateral-size effect described in this section. In other words, the lateral-size effect alone cannot account for the observation that $d_{c(film)} < 2$ *nm*.

## 7. A ferroelectric nanodot with nonferroelectric surface layers

According to the result of the previous section, the observed difference in the *w²-d* scaling behavior between a film and a nanodot [figure 3] cannot be explained by the critical-thickness model as quantified in equation (19). To resolve this puzzling problem, we postulate the presence of a ferroelectrically inactive surface layer with the thickness of $\delta_l$. This postulation is based on the experimental observations of a thin nonferroelectric surface layers in perovskite-based ferroelectrics such as BaTiO$_3$ [24-26]. According to the experimental estimate by transmission electron microscopy, the thickness of this surface layer is around 10 *nm* [25]. First-principles calculations [27,28] and phase-field simulations [29] also support the existence of a surface relaxation layer. In the case of first-principles studies [27,28], however, a substantially smaller value of the surface layer (~1 nm) is predicted. The size-dependent depolarization effects on $d_{33}$ and $E_c$, as shown in figure 1(d), also suggest the presence of a nonferroelectric surface layer. The most important effect of this postulation is considered to be the compensation of the depolarizing field [30] by the lateral surfaces of a nanodot.

Let us define $\delta_l$ as the thickness of the ferroelectrically inactive layer on the four side surfaces of a nano-rectangle and $\delta_t$ as the thickness of the nonferroelectric layer on the top surface. On the other hand, $\delta_b$ is defined as the thickness of the ferroelectrically inactive layer on the bottom surface which is compensated by free charge from the bottom conducting substrate, Nb-doped STO. Then, the Gibbs free-energy function $(\Delta G'^{dot})$ of a ferroelectric nanodot having nonferroelectric surface layers (with respect to a paraelectric nanodot of the same size) can be obtained by suitably modifying equation (1) under the presence of nonferroelectric surface layers, namely,

$$\Delta G'^{dot} = -\Delta \mu (L-2\delta_l)^2 \{d-(\delta_t+\delta_b)\} + 0*\left[L^2 d - (L-2\delta_l)^2 \{d-(\delta_t+\delta_b)\}\right]$$
$$+ \alpha P_o^2 (L-2\delta_l)^2 \{d-(\delta_t+\delta_b)\}\left(\frac{w}{\{d-(\delta_t+\delta_b)\}}\right) + \sigma_w \left\{\frac{(L-2\delta_l)}{w}-1\right\}(L-2\delta_l)\{d-(\delta_t+\delta_b)\}$$
$$+ \sigma_{fp}\left[4(L-2\delta_l)\{d-(\delta_t+\delta_b)\}+2(L-2\delta_l)^2\right] + \sigma_s'(4Ld+2L^2) \tag{20}$$

The second term in the right-hand side of equation (20) implies that the nonferroelectric surface layers do not contribute to the free energy as these layers can be regarded as paraelectric layers (reference phase). On the other hand, $\alpha P_o^2 w$, appeared in the third term indicates that the depolarizing-field energy in equation (20) is accurate for $d \geq d_c$. $\sigma_{fp}$ appeared in the fifth term denotes the interfacial tension between the ferroelectric nanodot and the paraelectric surface layer. The last term is practically zero as $\sigma_s' \equiv \sigma'(T,P) - \sigma_{pv} = \sigma_{pv} - \sigma_{pv} = 0$, where $\sigma_{pv}$ is the surface tension at the vapor-paraelectric-layer phase boundary This is because $\Delta G'^{dot}$ is expanded with respect to the free energy of the paraelectric nanodot having the same size.

The presence of the ferroelectrically inactive surface layer significantly reduces (i) the bulk free-energy difference between the paraelectric and ferroelectric phases ($\Delta \mu$) and (ii) the depolarizing-field energy [30] by a factor of $(L-2\delta_l)^2/L^2$ (geometry factor). Variations of $\Delta G'$ caused by other effects (i.e., domain-wall energy + ferro/para interfacial energy) are relatively less important. Considering these, one can establish the following approximate relation for the Gibbs free energy of a ferroelectric nanodot (per unit area):

$$\Delta G' \equiv \Delta G'^{dot} / L^2 \approx -\Delta \mu \frac{(L-2\delta_l)^2}{L^2}\{d-(\delta_t+\delta_b)\} + \alpha P_o^2 \frac{(L-2\delta_l)^2}{L^2} w$$
$$+ \sigma_w \left\{\frac{(L-2\delta_l)}{w}-1\right\}\frac{\{d-(\delta_t+\delta_b)\}}{(L-2\delta_l)} + 2\sigma_{fp}\left[\frac{2\{d-(\delta_t+\delta_b)\}}{(L-2\delta_l)}+1\right] \tag{21}$$

For simplicity, let us assume the following equality: $\delta_t = \delta_b = \delta_l \equiv \delta$. By adopting this equality and substituting the obvious requirement of $L - 2\delta_l = L - 2\delta = nw$ into equation (21), one obtains the following expression of $\Delta G'$ in terms of $nw$ and $\delta$:

$$\Delta G' \approx \frac{(nw)^2}{(nw+2\delta)^2}\left\{-\Delta\mu(d-2\delta) + \alpha P_o^2 w\right\} + \sigma_w \frac{(d-2\delta)}{w}\left(\frac{n-1}{n}\right) + 2\sigma_{fp}\left\{\frac{2(d-2\delta)}{nw}+1\right\}$$

$$\equiv g(w)\left\{-\Delta\mu(d-2\delta) + \alpha P_o^2 w\right\} + \sigma_w \frac{(d-2\delta)}{w}\left(\frac{n-1}{n}\right) + 2\sigma_{fp}\left\{\frac{2(d-2\delta)}{nw}+1\right\} \quad (22)$$

where $g(w)$ in the above equation denotes the geometry factor. Thus, we have

$$g(w) \equiv \frac{(L-2\delta)^2}{L^2} = \frac{(nw)^2}{(nw+2\delta)^2} \quad (23)$$

Then, the modified equilibrium domain width (in the presence of a ferroelectrically inactive surface layer) can be found by taking a partial derivative of $\Delta G'$ with respect to $w$, namely, $(\partial \Delta G'/\partial w)_d = 0$. However, direct partial differentiation of equation (22) leads to a very complex relation in which the equilibrium domain width ($w'_{eq}$) cannot be expressed analytically in terms of the relevant physical parameters that include $d, \delta, n,$ and $L$. This complication comes from $(\partial g(w)/\partial w)_\delta$. Thus, let us make some simplification to obtain a useful approximate relation for $w'_{eq}$. From equation (23), one obtains

$$(\partial g(w)/\partial w)_\delta = \frac{4n^2 w\delta}{(nw+2\delta)^3} < \frac{4n^2 w\delta}{(nw)^3} = \frac{4}{w}\left(\frac{\delta}{nw}\right) \approx 0 \quad (24)$$

The last approximate expression of equation (24) comes from the fact that $\delta << nw$. Thus, $g(w)$ in equation (22) can be treated as a constant when taking partial differentiation with respect to $w$. Then, using equation (22) and $(\partial \Delta G'/\partial w)_d = 0$, one obtains the following analytical expression for the modified equilibrium domain width in the presence of a nonferroelectric layer of the thickness $\delta$:

$$(w'_{eq})^2 = \left(\frac{d-2\delta}{\alpha P_o^2}\right)\left(\frac{L}{L-2\delta}\right)^2\left\{\left(\frac{n-1}{n}\right)\sigma_w + \frac{4\sigma_{fp}}{n}\right\} \quad (25)$$

It is interesting to note that the above equation reduces to equation (14) as $\delta \to 0$. Thus, equation (25) can be viewed as a modified LLK scaling law for a ferroelectric nanodot, which takes into account the presence of a nonferroelectric surface layer of the thickness $\delta$.

We are now in a position to examine the difference (or ratio) in $d_c$ between a film and a nanodot in the presence of a nonferroelectric surface layer. Again, we use the following proportionality: $w'_{c(dot)} = k\, d'_{c(dot)}$. By exploiting this proportionality and equation (25) in the vicinity of $d'_c$, one can eliminate $w'_{c(dot)}$ from this proportionality relation and obtain the following relation:

$$\frac{(d'_c)^2}{(d'_c - 2\delta)} = \left(\frac{1}{\alpha P_o^2 k^2}\right)\left(\frac{L_c}{L_c - 2\delta}\right)^2 \left\{\left(\frac{n_c - 1}{n_c}\right)\sigma_w + \frac{4\sigma_{fp}}{n_c}\right\} \tag{26}$$

The left-hand-side of equation (26) can be replaced by the following linearity approximation:

$$\frac{(d'_c)^2}{(d'_c - 2\delta)} = \left\{1 + \frac{2\delta}{(d'_c - 2\delta)}\right\} d'_c \approx d'_c + 2\delta \tag{27}$$

Using equation (27), one obtains the following linear approximation for the critical thickness.

$$\begin{aligned}
d'_{c(dot)} &= \frac{1}{\alpha P_o^2 k^2}\left(\frac{L_c}{L_c - 2\delta}\right)^2\left\{\left(\frac{n_c - 1}{n_c}\right)\sigma_w + \frac{4\sigma_{fp}}{n_c}\right\} - 2\delta \\
&= \frac{5\pi\varepsilon_o}{P_o^2}\left(\frac{\varepsilon_z^{3/2}}{\varepsilon_x^{1/2}}\right)\left(\frac{L_c}{L_c - 2\delta}\right)^2\left\{\left(\frac{n_c - 1}{n_c}\right)\sigma_w + \frac{4\sigma_{fp}}{n_c}\right\} - 2\delta \approx \frac{5\pi\varepsilon_o \sigma_w}{P_o^2}\left(\frac{\varepsilon_z^{3/2}}{\varepsilon_x^{1/2}}\right)
\end{aligned} \tag{28}$$

where $k^2$-term appeared in the first expression was eliminated in the second expression by comparing equation (11) with the first expression in the asymptotic limit of $n \to \infty$ and $\delta \to 0$. Thus, the last expression of equation (28) corresponds to the asymptotic thin-film limit ($= d_{c(film)}$) and does coincide exactly with equation (11).

Accordingly, we have the following ratio for the critical thickness in the presence of a non-ferroelectric surface layer with the thickness $\delta$:

$$R_\delta \equiv \frac{d'_{c(dot)}}{d'_{c(film)}} = \left(\frac{L_c}{L_c - 2\delta}\right)^2 \left\{\left(\frac{n_c - 1}{n_c}\right) + \frac{4\sigma_{fp}}{n_c \sigma_w}\right\} \geq 1 \quad (29)$$

where $R_\delta$ denotes the ratio of the two critical thicknesses in the presence of a nonferroelectric surface layer of the thickness $\delta$. The most prominent difference between equation (19) and equation (29) is the introduction of a geometry term, $\{L_c/(L_c - 2\delta)\}^2$ in equation (29), which always enhances $R_\delta$. More importantly, equation (29) tells us that both the existence of a nonferroelectric layer of the thickness $\delta$ and the finite lateral-size effect determine $R_\delta$, thus $d'_{c(dot)}$. Taking $(\sigma_{fp}/\sigma_w) \approx 1/2$ and plugging $n_c = 3$, $L_c \approx 45$ $nm$, and $\delta_l \approx 17$ $nm$ into equation (29), one obtains $d'_{c(film)}$ of 1.57 $nm$ (< 2 $nm$), which qualitatively explains the observed validity of the LLK scaling law down to ~2 $nm$ [5,14,15]. We thus conclude that the existence of a nonferroelectric surface layer, in addition to the lateral-size effect ($n$), should be taken into account to properly explain the difference in the scaling behavior between ferroelectric films and nanodots.

## 8. Conclusions

In conclusion, for the dot thickness larger than the critical value ($d_c$) of ~35 nm, the width of 180° stripe domains scales with the thickness according to the LLK scaling law. For the dot thickness smaller than $d_c$, however, we obtain a quite striking correlation that the thickness-dependent domain width is characterized by a negative exponent (*i.e.*, $\gamma < 0$). On the basis of theoretical considerations of $d_c$, we attribute this anomalous domain periodicity to the existence of a nonferroelectric surface layer, in addition to the finite lateral-size effect of a ferroelectric nanodot.

**Acknowledgment**

This work was supported by the Brain Korea 21 project 2012 and by the World Class University (WCU) program through the National Research Foundation (NRF) funded by the Ministry of Education, Science and Technology of Korea (Grant No. R31-2008-000-10059-0).

**References**

[1] Scott J F 2000 *Ferroelectric Memories* (Springer-Verlag, Berlin)

[2] Mitsui T and Furuichi J 1953 *Phys. Rev.* **90** 193-202

[3] Shih W Y, Shih W-H and Aksay I A 1994 *Phys. Rev. B* **50** 15575-15585

[4] Bratkovsky A M and Levanyuk A P 2000 *Phys. Rev. Lett.* **84** 3177-3180

[5] Streiffer S K, Eastman J A, Fong D D, Thompson C, Munkholm A, Ramana Murty M V, Auciello O, Bai G R and Stephenson G B 2002 *Phys. Rev. Lett.* **89** 067601

[6] Fong D D, Stephenson G B, Streiffer S K, Eastman J A, Auciello O, Fuoss P H and Thompson C 2004 *Science* **304**, 1650-1653

[7] Lai B-K, Ponomareva I, Naumov I I, Kornev I, Fu H, Bellaiche L and Salamo G J 2006 *Phys. Rev. Lett.* **96**, 137602

[8] Naumov I and Bratkovsky A M 2008 *Phys. Rev. Lett.* **101**, 107601

[9] Schilling A, Bryne D, Catalan G, Webber K G, Genenko Y A, Wu G S and Scott J F 2009 *Nano Lett.* **9**, 3359-3364

[10] Catalan G, Béa H, Fusil S, Bibes M, Paruch P, Barthélémy A and Scott, J F 2008 *Phys. Rev. Lett.* **100**, 027602

[11] Kittel C 1946 *Phys. Rev.* **70**, 965-971

[12] Schilling A, Adams T B, Bowman R M, Gregg J M, Catalan G and Scott J F 2006 *Phys. Rev. B* **74**, 024115

[13] Scott J F 2006 *J. Phys.: Condens. Matter* **18**, R361-R386

[14] Catalan G, Scott J F, Schilling A and Gregg J M 2007 *J. Phys.: Condens. Matter* **19**, 022201

[15] Zhao G.-P, Chen L and Wang J 2009 *J. Appl. Phys.* **105**, 061601

[16] Prosandeev S, Lisenkov S and Bellaiche L 2010 *Phys. Rev. Lett.* **105**, 147603

[17] Burns G and Scott B A 1970 *Phys. Rev. Lett.* **25**, 167-170

[18] Cho S M, Jang H M and Kim T Y 2001 *Phys. Rev. B* **64**, 014103

[19] Scott J F 2007 *Science* **315**, 954-959

[20] Son J Y, Shin Y-H, Ryu S, Kim H and Jang H M 2009 *J. Am. Chem. Soc.* **131**, 14676-14678

[21] Kopal A, Bahnik T and Fousek J 1997 *Ferroelectrics* **202**, 267-274

[22] Mitsui T 1976 *An Introduction to the Physics of Ferroelectrics* (Gordon and Breach Publishers, London). pp. 59-77

[23] Bjorkstam J L and Oettel R E 1967 *Phys. Rev.* **159**, 427-430

[24] Känzig W 1955 *Phys. Rev.* **98**, 549-550

[25] Tsai F and Cowley J M 1994 *Appl. Phys. Lett.* **65**, 1906-1908

[26] Kolpak A M, Li D, Shao R, Rappe A M and Bonnell D A 2008 *Phys. Rev. Lett.* **101**, 036102

[27] Padilla J and Vanderbilt D 1997 *Phys. Rev. B* **56**, 1625-1631

[28] Bungaro C and Rabe K M 2005 *Phys. Rev. B* **71**, 035420

[29] Slutsker J, Artemev A and Roytburd A 2008 *Phys. Rev. Lett.* **100**, 087602

[30] Fong D D, Kolpak A M, Eastman J A, Streiffer S K, Fuoss, P H, Stephenson G B, Thompson C, Kim D M, Choi K J, Eom C B, Grinberg I and Rappe A M 2006 *Phys. Rev. Lett.* **96**, 127601

# * Figure Captions

**Figure 1.** (a) Dip-pen nanolithography of PbTiO$_3$ nanodots on a Nb-doped SrTiO$_3$ substrate. (b) An AFM image for PbTiO$_3$ nanodot array having four different size classes. The array was formed by a dip-pen nanolithography method. (c) Side length and thickness of PbTiO$_3$ nanodots as a function of the dip-pen deposition time. (d) $d_{33}$ value and the coercive electric field ($E_c$) plotted as a function of the nanodot size, where A, B, C, and D indicate the dot size presented in (b).

**Figure 2.** (a) PFM images of the five selected PbTiO$_3$ nanodots with different lateral sizes. (b) A PFM line profile of the PbTiO$_3$ nanodot having a lateral dimension of 185 *nm* as an example. (c) AFM line profiles of the four selected PbTiO$_3$ nanodots with different lateral sizes, where A, B, C, and D denote the lateral size (i.e., side length) of 45, 60, 100, and 150 *nm*, respectively.

**Figure 3.** The thickness-dependent domain periodicity (*w*) of the PbTiO$_3$ nanodot is compared with that of the epitaxially grown PbTiO$_3$ film. Data for the thin films (blue colour triangles) were taken from *Phys. Rev. Lett.,* **89** (6), 067601 (2002).

**Figure 4.** A schematic representation of 180$^o$ stripe domains in a given nanodot, where *L*, *w*, and *d* are described in the text.

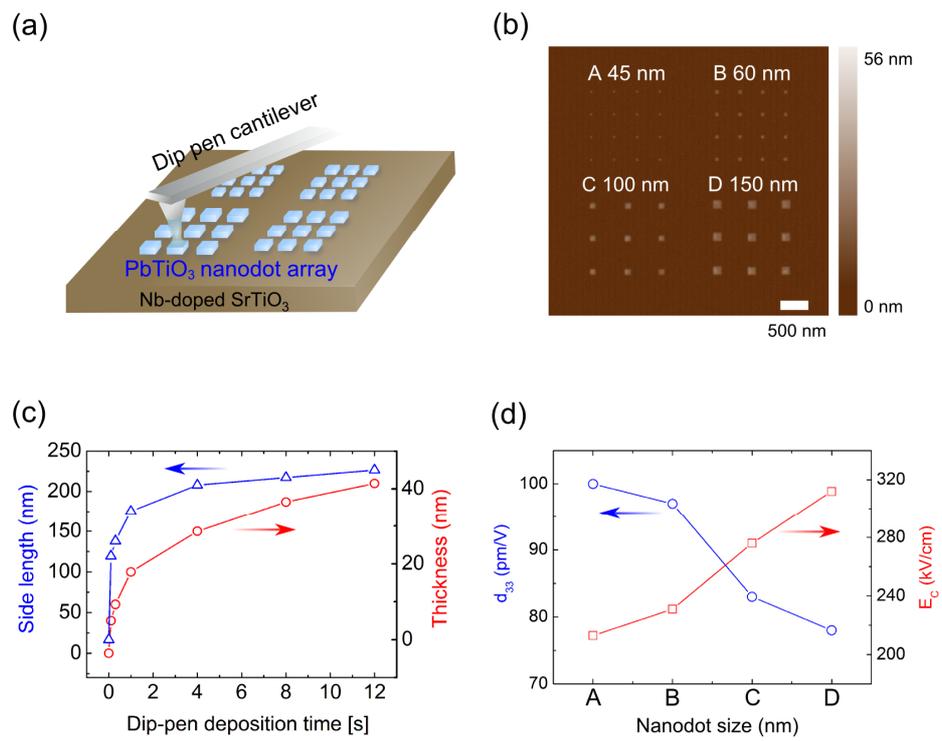

**Figure 1**
21

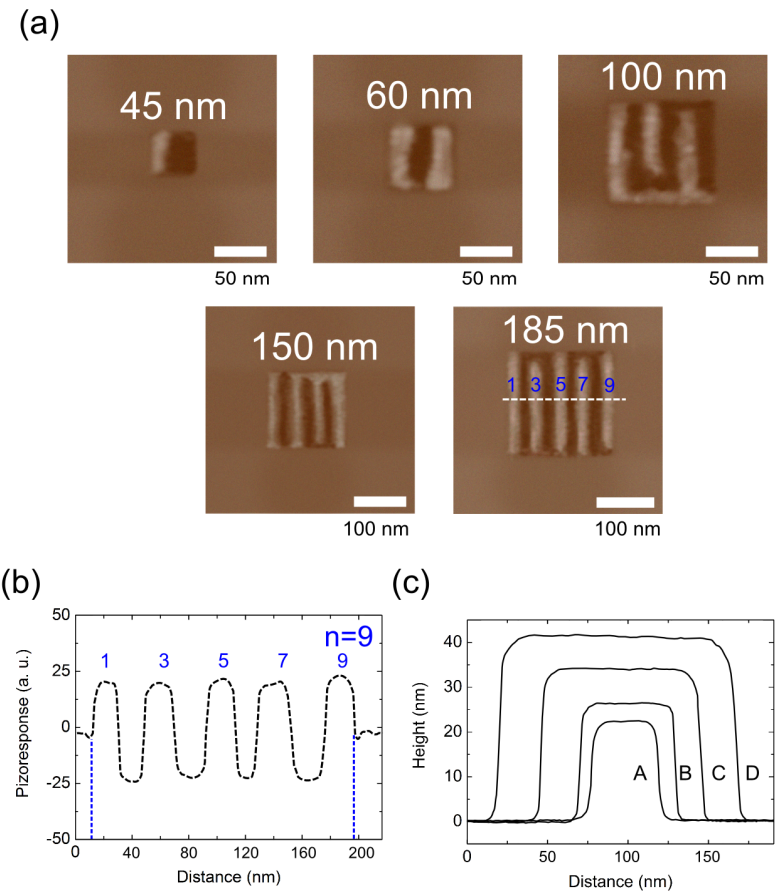

**Figure 2**

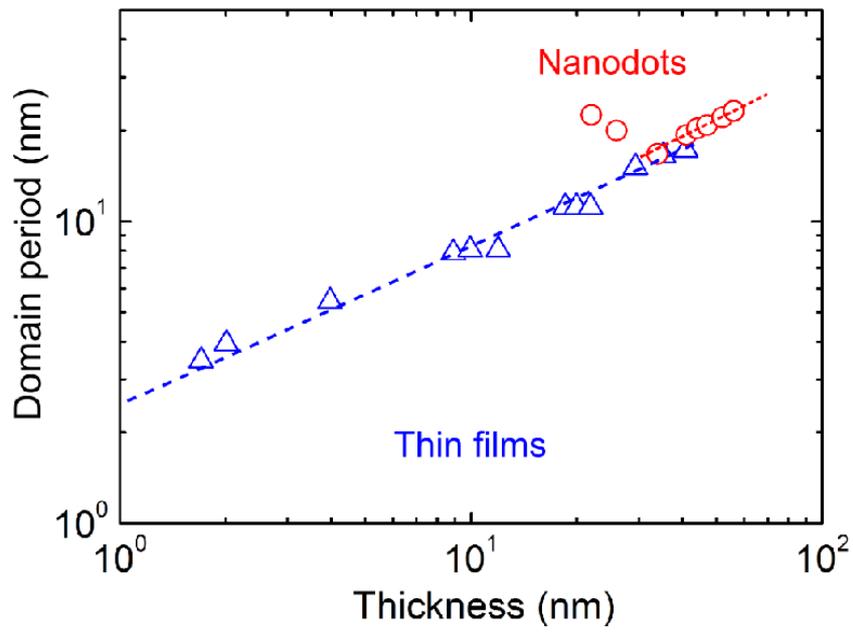

**Figure 3**

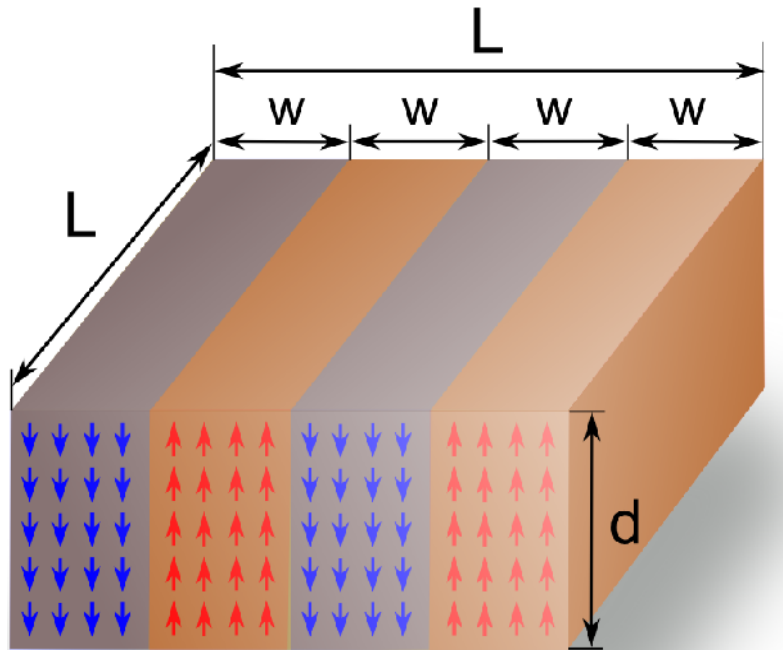

**Figure 4**